\RequirePackage[T1]{fontenc}
\documentclass[12pt]{article}
\usepackage[height=8.85in,width=6.45in]{geometry}

\usepackage[utf8]{inputenc}

\usepackage{amsmath}
\usepackage{amssymb}
\usepackage{mathtools}
\usepackage{amsthm}%theorem environment

\usepackage{times}
\usepackage[scaled]{couriers}
\usepackage{mathrsfs}

\usepackage{graphicx}%includegraphics
\usepackage{subcaption}%subcaption

\usepackage{tikz}
\usepackage{tikz-cd}
\usetikzlibrary{calc}

\usepackage[svgnames]{xcolor}
\usepackage[colorlinks,linktocpage=true,citecolor=DarkGreen,linkcolor=FireBrick]{hyperref}
\usepackage{cite}

\usepackage{bm}
\usepackage{slashed}
\usepackage{braket}

\numberwithin{equation}{section}

\theoremstyle{plain}
\newtheorem{thm}{Theorem}

\theoremstyle{definition}

\numberwithin{thm}{section}

\def\i{{\mathsf i}}

\def\Hom{\mathop{\mathrm{Hom}}}

\def\bZ{{\mathbb Z}}

\def\U{\mathrm{U}}

\def\Pin{\mathrm{Pin}}
\def\SL{\mathrm{SL}}
\def\Mp{\mathrm{Mp}}
\def\GL{\mathrm{GL}}

\usepackage{amsmath}
\usepackage{mathtools}

\begin{document}

\begin{titlepage}

\begin{flushright}
OU-HET-1289\\
KOBE-COSMO-25-14 \\
TU-1274
\end{flushright}

\vskip 3cm

\begin{center}

{\large \bfseries Emergent higher-form symmetry from type~IIB superstring theory}

\vskip 1cm
Naoto~Kan$^1$, Masashi~Kawahira$^2$, and Hiroki~Wada$^3$
\vskip 1cm

\begin{tabular}{ll}
$^1$ &Department of Physics, The University of Osaka,
Toyonaka 560-0043, Japan \\
$^2$ & Department of Physics, Kobe University, Kobe 657-8501, Japan\\
$^3$ & Department of Physics, Tohoku University, Sendai 980-8578, Japan\\
\end{tabular}

\vskip 1cm

\end{center}

\noindent
We investigate a higher-form symmetry in type IIB superstring theory, which possesses an $\SL(2,\bZ)$ symmetry. 
From the point of view of the low-energy effective field theory, the $\SL(2,\bZ)$ symmetry is treated as a gauge symmetry.
Hence, an $8$-form global symmetry $\bZ_{12}^{[8]}$ emerges as a quantum symmetry.
In this paper, we present an explicit construction of the topological operator associated with the $\bZ_{12}^{[8]}$ symmetry.
In this construction, the discriminant $\Delta(\tau)$ plays a central role.
As a result, it becomes manifest that $\bZ_{12}^{[8]}$ is the solitonic symmetry of $7$-branes.
Furthermore, taking into account the extensions of the duality group, we  also discuss what global symmetries emerge when considering not $\SL(2,\bZ)$ but $\Mp(2,\bZ)$, $\GL(2,\bZ)$, and $\Pin^+(\GL(2,\bZ))=\GL^+(2,\bZ)$.

\end{titlepage}

\setcounter{tocdepth}{3}
%\tableofcontents

%\newpage

\tableofcontents

\newpage

\section{Introduction and summary}

Superstring theory is a remarkable framework. It is one of the few theories that is UV-complete~\cite{Scherk:1974ca} and allows for a consistent computation of graviton scattering~\cite{Veneziano:1968yb}. Moreover, it tells us about non-perturbative effects in quantum field theory, such as dualities.

Superstring theory is one of the candidates for the ultimate theory of our universe, and several efforts are made to derive observational or phenomenological predictions from it.
However, there is a significant difference between superstring theory and the world we observe.
It is plausible that superstring theory admits no global symmetries~\cite{Banks:1988yz,Banks:2010zn}, while our world exhibits them.
If superstring theory indeed describes reality, then the symmetries we observe must be approximate in some sense.
Global symmetries that are absent in a microscopic theory but appear at a macroscopic scale are called \textit{emergent symmetries}.
Such global symmetries, unlike gauge symmetries, are invariant under renormalization group transformations and therefore impose strict constraints on the macroscopic theory~\cite{tHooft:1979rat}.

It is thus an interesting problem to understand what emergent symmetries can arise in the low-energy effective theories of superstring theory.
In particular, this work focuses on the emergent symmetry originating from type IIB superstring theory.
This theory enjoys $\SL(2,\bZ)$ symmetry which enables us to describe the theory non-perturbatively.
This framework is known as \textit{F-theory}~\cite{Vafa:1996xn}.
F-theory is not only mathematically intriguing but has also been extensively discussed in connection with the building of phenomenological models such as grand unified theories (GUTs) ~\cite{Donagi:2008ca,Beasley:2008dc,Beasley:2008kw,Donagi:2008kj}.

In this paper, we investigate the emergent symmetry by combining the F-theoretic description with the notion of generalized symmetry~\cite{Kapustin:2014gua,Gaiotto:2014kfa}.
\footnote{
There are several works on the relation between higher-form symmetry and string theory, for instance, \cite{DelZotto:2015isa,Morrison:2020ool,Albertini:2020mdx,DelZotto:2020esg,Bhardwaj:2020phs,Gukov:2020btk,Heidenreich:2020pkc,Apruzzi:2021nmk,Bhardwaj:2021mzl,Genolini:2022mpi,DelZotto:2022fnw,Cvetic:2022imb,Apruzzi:2022dlm,Hubner:2022kxr,Heckman:2022muc,Grimm:2022xmj,Etheredge:2023ler,Amariti:2023hev,Cvetic:2023pgm,Lawrie:2023uiu,DelZotto:2024tae,Braeger:2024jcj,Franco:2024mxa,Baume:2024oqn,Cvetic:2024mtt,Tian:2024dgl,Braeger:2025rov,Najjar:2024vmm,Najjar:2025htp,Khlaif:2025jnx}.
}
By employing the framework of generalized symmetries, one finds that in local quantum field theory it is not always possible to gauge all symmetries simultaneously.
In general, when one symmetry is gauged, another global symmetry inevitably emerges.
This global symmetry is called the \textit{quantum symmetry}~\cite{Vafa:1989ih,Gaiotto:2014kfa}.
By making use of this quantum symmetry, we identify the emergent symmetry that arises in the low-energy limit of superstring theory.

Our work can be summarized as follows.
We first discuss the quantum symmetry of a $p$-form symmetry $G^{[p]}$. 
Generally, 
if we gauge a $p$-form symmetry $G^{[p]}$, the quantum symmetry is $(d-p-2)$-form symmetry $\widehat{G}^{[d-p-2]}$, where $\widehat{G}=\Hom(G,\U(1))$. 
In addition to the case that $G^{[p]}$ is abelian, we also consider the non-abelian case.\footnote{ 
If $G$ is non-abelian, then since there are no non-abelian higher-form symmetries, $p$ must be zero.
}
It turns out that the quantum symmetry of a non-abelian discrete symmetry can be thought as that of the abelianization. 
Thus, the $(d-2)$-form quantum symmetry $\widehat{G}^{[d-2]}$ is characterized by $\widehat{G}=\Hom(G^{\rm ab},\U(1))$, where ${G}^{\rm ab}$ is the abelianization of the non-abelian group $G$.
Through this analysis, it becomes clear 
why the abelianization plays an important role in F-theory.

Furthermore, based on this discussion, we find that type IIB supergravity possesses an emergent symmetry  
$\bZ_{12}^{[8]}$
originating from the 
$\SL(2,\bZ)$ gauge symmetry.\footnote{
In the context of the swampland cobordism conjecture, the significance of the $\mathbb{Z}_{12}$ group is also discussed~\cite{Dierigl:2020lai}.
However, it should rather be interpreted as a $\mathbb{Z}_{12}$ gauge symmetry induced from the $\SL(2,\bZ)$ gauge symmetry, and therefore it exists as an exact symmetry in string theory itself.
In contrast, in this paper, we discuss a quantum symmetry --- a global symmetry that exists only effectively in the low-energy regime. In fact, this global symmetry is explicitly broken in quantum gravity scale. See also Footnote \ref{footnote:dynamical_7_brane}.
\label{footnote:swampland}}
We explicitly construct the topological operator associated with this symmetry.
The topological operator of $\bZ_{12}^{[8]}$ supported on the closed loop $\gamma$ is given by
\begin{align}
\begin{aligned}
&{\cal W}_\phi(\gamma)=\exp\left({\rm i} \phi {\cal Q}(\gamma)\right),& 
&\mathcal{Q}(\gamma)=\frac{1}{2\pi {\rm i}}\oint_\gamma {\rm d}z\frac{{\rm d}}{{\rm d}z}\log\Delta(\tau(z)),
\end{aligned}
\end{align}
where $\Delta(\tau(z))$ is the discriminant of the Weierstrass form, and $\tau(z)$ is the axio-dilaton in type IIB supergravity.
It is worth noting that, despite the discrete nature of this emergent symmetry, one can explicitly construct its charge operator using the local field $\tau(z)$.
Furthermore, by focusing on the extensions of the duality group, 
$\SL(2,\bZ)$ is replaced by the $\Pin^+$ version of the double cover of $\GL(2,\bZ)$, i.e., $\GL^+(2,\bZ)$.
As a consequence, one finds that the emergent symmetry $\bZ_{24}^{[8]}\rtimes \bZ^{[0]}_2$ arises.

Finally, 
we comment on subtle issues in extracting low-energy effective quantum field theories from superstring theory.
Naively, by taking the string length to zero, $\ell_{\rm string}\to 0$, one expects to obtain a field theory containing only massless degrees of freedom.
However, there remains freedom of choice regarding the heavy degrees of freedom (topological phases), leading to subtle distinctions, which we discuss.

The rest of this paper is organized as follows. 
In Section \ref{sec:Brief review of quantum symmetry}, we first briefly review the quantum symmetry.
In Section \ref{sec:Quantum symmetry of SL(2,Z)}, we construct the topological operator associated with the emergent symmetry in type IIB supergravity.
In Section \ref{sec:Extensions of the duality group and emergent symmetries}, we discuss the emergent symmetries that arise when considering the extensions of the duality group.
Section \ref{sec:Discussions} is devoted to discussions.

\section{Brief review of quantum symmetry}
\label{sec:Brief review of quantum symmetry}

In this section, we recall quantum symmetries~\cite{Vafa:1989ih,Gaiotto:2014kfa}.
Starting with a discrete abelian group, we generalize the discussion to arbitrary discrete group, including non-abelian cases.

\subsection{Abelian case}
\label{subsec:abelian case}

We begin with the case of a discrete abelian group.
Let $\mathcal{T}$ be a theory with a $p$-form discrete abelian global symmetry $G^{[p]}$.
Then the theory $\mathcal{T}/G^{[p]}$, obtained by gauging $G^{[p]}$, possesses a $(d-p-2)$-form global symmetry $\widehat{G}^{[d-p-2]}$.
Here, $\widehat{G}$ denotes the Pontryagin dual of $G$, i.e., $\Hom(G,\U(1))$.

We can see the emergence of the $(d-p-2)$-form symmetry as follows:
In $\mathcal{T}/G^{[p]}$, there exists a $(p+1)$-form gauge field $a^{(p+1)}$, which is the gauge field of the gauged $G^{[p]}$ symmetry.
We define its Wilson operator as
\begin{align}
    W_\phi(\Sigma^{(p+1)})
    =
    \exp\left(\i\phi\oint_{\Sigma^{(p+1)}} a^{(p+1)}\right).
\end{align}
By definition, this operator is gauge invariant.
Since $G^{[p]}$ is a discrete symmetry, $W_\phi(\Sigma^{(p+1)})$ is a topological operator.
Noting that $\phi\in \Hom(G,\U(1))$, we conclude that there exists $(d-p-2)$-form global symmetry $\widehat{G}^{[d-p-2]}$.

We refer to $\widehat{G}^{[d-p-2]}$ as the \textit{quantum symmetry} or the {\it dual symmetry}.
We note that since $\, \widehat{\!\widehat{G}}=G$, the quantum symmetry of the quantum symmetry coincides with the original symmetry.
In many cases, for $\mathcal{T}'=\mathcal{T}/G^{[p]}$, by further gauging, one can \textit{recover} the original theory. That is,
\begin{align}
\mathcal{T}'/\widehat{G}^{[d-p-2]}=\mathcal{T}. 
\end{align}

\subsection{Example: compact scalar theory}
\label{subsec:Example : compact scalar theory}
Let us see a simple example of the quantum symmetry for an abelian discrete group.
Consider the $d$-dimensional massless scalar theory:
\begin{align}
\mathcal{L}=\frac{1}{2}(\partial_\mu\Phi)^2.
\end{align}
This theory possesses a $0$-form global symmetry $\mathbb{R}^{[0]}$ that acts on the scalar field $\Phi$ as
\begin{align}
\Phi(x)\mapsto \Phi(x)+\alpha f\ \ \ (\alpha\in\mathbb{R}),
\end{align}
where $f$ is a constant that has an appropriate mass dimension.

Let us now consider gauging the subgroup symmetry $\bZ^{[0]}\subset \mathbb{R}^{[0]}$.
Based on the discussion in Section~\ref{subsec:abelian case}, since $\Hom(\bZ,\U(1))\cong \U(1)$
% \begin{align}
% \Hom(\bZ,\U(1))\cong \U(1),
% \end{align}
the gauged theory should have a $(d-2)$-form global symmetry $\U(1)^{[d-2]}$.
In fact, gauging $\bZ^{[0]}$ is equivalent to imposing the identification
\begin{align}
\Phi(x)\sim \Phi(x)+f.
\end{align}
Therefore $\Phi$ becomes a compact scalar with a decay constant $f$ after gauging.
As a result, we can define the following topological operator supported on a closed loop $\gamma$  as
\begin{align}
W_\phi(\gamma)
=
\exp\left(\frac{{\rm i}\phi}{f}\oint_\gamma {\rm d}\Phi\right).
\end{align}
Since $\frac{1}{f}\oint_\gamma {\rm d}\Phi\in\bZ$, 
we see that $\phi\in \U(1)$.
Thus, this operator generates the $(d-2)$-form global symmetry $\U(1)^{[d-2]}$.

There is another way to understand the above discussion.
Recall that $B\bZ\cong S^1$.
Let $M$ be the spacetime. Then a $\bZ$ gauge field,
\begin{align}
a:M\to B\bZ,
\end{align}
can be interpreted as a compact scalar field,
\begin{align}
\Phi:M\to S^1.
\end{align}
Therefore, the topological operator $W_\phi(\gamma)$ can be regarded as the Wilson operator of the $\bZ$ gauge field.

\subsection{General case}
\label{subsec:General case}
Let us now consider the quantum symmetry of an arbitrary discrete group.
At this point, we have the following important theorem:

\begin{thm}
Let $G$ be an arbitrary group and $A$ be an abelian group, and consider a homomorphism $\phi:G\to A$. 
Let $G^{\rm ab}$ denotes the abelianization of $G$. 
Then there exists a homomorphism $\varphi:G^{\rm ab}\to A$ such that the following diagram commutes:
\[
\begin{tikzcd}
G \arrow[rr,"\phi"] \arrow[dr,"{\rm proj}"'] && A \\
& G^{\rm ab} \arrow[ur,"\varphi"'] &
\end{tikzcd}
\]
In the commutative diagram above, ${\rm proj}:G\to G^{\rm ab}$ is the canonical projection defined by $G^{\rm ab}:=G/[G,G]$ where $[G,G]$ denotes the commutator subgroup.
\end{thm}

\noindent
This theorem implies that
\begin{align}
\Hom(G,\U(1))
\cong 
\Hom(G^{\rm ab},\U(1)).
\label{eq:Hom_G_U(1)_Hom_G_ab_U(1)}
\end{align}
Physically, a Wilson operator constructed from the $G$ gauge field and labeled $\phi\in\Hom(G,\U(1))$ can be interpreted as a Wilson operator associated with the $G^{\rm ab}$ gauge field, labeled by $\varphi\in\Hom(G^{\rm ab},\U(1))$.
Thus when $G$ is a discrete non-abelian group, the quantum symmetry is characterized by its abelianization $G^{\rm ab}$.
Namely, the theory $\mathcal{T}/G^{[p]}$, obtained by gauging $G^{[p]}$, possesses a $(d-p-2)$-form global symmetry $\widehat{G}^{[d-p-2]}$, where $\widehat{G}:=\Hom(G^{\rm ab},\U(1))$.
We apply the discussion here to the ${\rm SL}(2,\bZ)$ symmetry in type IIB superstring theory in the next section.

For a non-abelian group ${G}$, the identity $\, \widehat{\!\widehat{G}}=G$ does not hold since 
\begin{align}
\widehat{\!\widehat{G}}=\widehat{\widehat{G^{\rm ab}}}=G^{\rm ab}.  
\label{eq:G_hat_hat=G_abel}
\end{align}
We interpret that the quantum symmetry of the quantum symmetry is identified with the abelianization of the original symmetry.
Therefore, in $\left(\mathcal{T}/G^{[p]}\right)/\hat{G}^{[d-p-2]}$ obtained by gauging twice,
it, in general, differs from the original theory $\mathcal{T}$,
because 
these theories have distinct symmetries.

To recover the original theory $\mathcal{T}$ from $\mathcal{T}/G^{[p]}$, one needs to take into account that $\mathcal{T}/G^{[p]}$ possesses a non-invertible symmetry~\cite{Bhardwaj:2017xup}.
In fact, $\mathcal{T}/G^{[p]}$ contains topological operators labeled by ${\rm Rep}(G)$.
After including these, one can recover $\mathcal{T}$.

\section{Quantum symmetry of \texorpdfstring{$\SL(2,\bZ)$}{SL(2,Z)} in type~IIB superstring}
\label{sec:Quantum symmetry of SL(2,Z)}

In this section, we consider type~IIB supergravity~\cite{Green:1982tk}, which is the low-energy effective theory of type~IIB superstring theory.
From the discussion in Section~\ref{sec:Brief review of quantum symmetry},
we see that this theory possesses a global $8$-form symmetry.

Since superstring theory itself does not admit any global symmetry~\cite{Banks:1988yz,Banks:2010zn}, this $8$-form symmetry should be interpreted as an \textit{emergent symmetry}.

\subsection{\texorpdfstring{$\SL(2,\bZ)$}{SL(2,Z)} gauge symmetry in type~IIB supergravity}

Type IIB supergravity contains the following fields:
the RR $0$-form $C^{(0)}$, the RR $2$-form $C^{(2)}_{\mu\nu}$ and the RR $4$-form $C^{(4)}_{\mu\nu\rho\sigma}$ in the RR sector;
the graviton $g_{\mu\nu}$, the Kalb–Ramond field $B^{(2)}_{\mu\nu}$ and the dilaton $\phi$ in the NSNS sector;
and %from the R–NS and NS–R sectors, 
the fermions in the RNS and NSR sectors.

 Type IIB supergravity has the $\SL(2,\mathbb{R})$ invariance, that is, let
\begin{align}
\begin{pmatrix}
   a & b \\
   c & d
\end{pmatrix}
\end{align}
be an element of $\SL(2,\mathbb{R})$, then
the action is invariant under the following transformations:\footnote{Here we omit the transformations of the fermionic fields.}
\begin{align}
    \tau(x)
\mapsto \frac{a\tau(x)+b}{c\tau(x)+d},\qquad
\begin{pmatrix}
   C^{(2)}_{\mu\nu}(x)  \\
   B^{(2)}_{\mu\nu}(x) 
\end{pmatrix}
\mapsto
\begin{pmatrix}
   a & b \\
   c & d
\end{pmatrix}
\begin{pmatrix}
   C^{(2)}_{\mu\nu}(x)  \\
   B^{(2)}_{\mu\nu}(x) 
\end{pmatrix},
\end{align}
where we have defined
\begin{align}
\tau(x)=C^{(0)}(x)+{\rm i}e^{-\phi(x)},
\end{align}
which is called the axio-dilaton.
The RR $4$-form field and the graviton are invariant under the transformation.

Type~IIB supergravity possesses an equivalence relation under 
$\SL(2,\bZ) \subset \SL(2,\mathbb{R})$.
In other words, the $\SL(2,\bZ)$ symmetry is gauged~\cite{Hull:1994ys}.\footnote{
Anomaly cancellation in type IIB theory with gauged $\SL(2,\bZ)$ is discussed in \cite{Debray:2021vob}.
}
In some references, the $\SL(2,\bZ)$ symmetry is referred to as a duality.
However, $\tau(x)$ is not a coupling constant but a field.
Therefore, strictly speaking, it is a symmetry.

\subsection{Emergent symmetry associated with D7-brane}

Let us now focus on the subgroup $\bZ\subset \SL(2,\bZ)$ embedded as
\begin{align}
\begin{array}{ccc}
\bZ
&
\hookrightarrow
&
\SL(2,\bZ)\\
\rotatebox{90}{$\in$}&& \rotatebox{90}{$\in$} \\
m & \longmapsto& 
\begin{pmatrix}
   1 & m \\
    & 1
\end{pmatrix}.
\end{array}
\end{align}
In this section, 
we assume that only the $\bZ$ symmetry is gauged,
that is, the identification,
\begin{align}
\tau(x)
\sim
\tau(x)
+
m
\end{align}
or
\begin{align}
C^{(0)}(x)
\sim
C^{(0)}(x)
+
m,
\end{align}
is imposed.
This implies that $C^{(0)}$ can be understood as a compact scalar field.

By the discussion in Section \ref{subsec:Example : compact scalar theory}, 
this theory has an $8$-form global symmetry $\U(1)^{[8]}$ whose topological operator is
\begin{align}\begin{aligned}\label{eq:q=oint_d_C_0}
W_\phi(\gamma)&=\exp\left({\rm i}\phi Q(\gamma)\right),& Q(\gamma)&:=\oint_\gamma {\rm d}C^{(0)}.
\end{aligned}\end{align}
The physical meaning of this topological operator is as follows:
the operator $Q(\gamma)$ represents the number of D7-branes enclosed by $\gamma$.
In other words, this emergent symmetry can be regarded as a solitonic symmetry~\cite{Mermin:1979zz,Chen:2022cyw} associated with the D7-branes.
However, this interpretation is valid when only the $\bZ$ subgroup is gauged.
If the full symmetry $\SL(2,\bZ)$ is gauged, then this topological operator is no longer gauge invariant.

\subsection{Emergent symmetry associated with 7-brane}
\label{subsec:Emergent symmetry associated with 7-brane}

Let us now consider the quantum symmetry of the full gauge symmetry $\SL(2,\bZ)$.\footnote{For some basic properties of the $\SL(2,\bZ)$ group, see e.g.~\cite{Seiberg:2018ntt}.}
Since $\SL(2,\bZ)$ is non-abelian, it is useful to see its abelianization, as discussed in Section \ref{subsec:General case}.

\begin{thm}
The abelianization of $\SL(2,\bZ)$ 
is $\bZ_{12}$.
\end{thm}
\noindent
By this theorem and Eq.~\eqref{eq:Hom_G_U(1)_Hom_G_ab_U(1)},
\begin{align}
\Hom(\SL(2,\bZ),\U(1))
\cong
\Hom(\bZ_{12},\U(1))
\cong
\bZ_{12}.
\end{align}
Then type IIB supergravity should have an $8$-form global symmetry $\bZ_{12}^{[8]}$.\footnote{
As mentioned in the outset of Section~\ref{sec:Quantum symmetry of SL(2,Z)}, the $8$-form global symmetry is an emergent symmetry of type~IIB supergravity. In type~IIB superstring theory, this symmetry is broken by \textit{dynamical} $7$-branes~\cite{Heidenreich:2020pkc}. See also~\cite{McNamara:2021cuo} for related discussions.\label{footnote:dynamical_7_brane}
}
\footnote{In the context of the Swampland cobordism conjecture, the significance of the $\mathbb{Z}_{12}$ group is also discussed. See also Footnote~\ref{footnote:swampland}.}
To construct a topological operator associated with the $8$-form symmetry explicitly, it is convenient to use an \textit{F-theory 
framework
}~\cite{Vafa:1996xn}.\footnote{See e. g.~\cite{Taylor:2011wt,Weigand:2018rez} for the review of F-theory.}

Let us consider a system on $\Sigma$ with 
parallel 7-branes, where $\Sigma$ denotes the two-dimensional surface transverse to the 7-branes.
Let $z$ be a complex coordinate of $\Sigma$.
Since the equivalence relation,
\begin{align}\begin{aligned}
    \tau(z)&\sim\frac{a\tau(z)+b}{c\tau(z)+d},&
\begin{pmatrix}
   a & b \\
   c & d
\end{pmatrix}
&\in\SL(2,\bZ),
\end{aligned}\end{align}
is imposed for the $\SL(2,\bZ)$ gauge symmetry, one can think that $\tau(z)$ determines a torus fibration over $\Sigma$.\footnote{
Here we assume that $\tau(z)$ is a holomorphic function, which is ensured by supersymmetry. If it is broken in the low energy regime, this assumption is no longer valid. Hence the explicit expression of the $\mathbb{Z}_{12}^{[8]}$ topological operator is unclear. Even in this case, the existence of $\mathbb{Z}_{12}^{[8]}$ is guaranteed by the discussion in Section \ref{sec:Brief review of quantum symmetry}.
}
It can be described by an elliptic fibration.
This is because $\tau$ defines an elliptic curve through the Weierstrass equation
\begin{align}
  y^2=x^3+f(\tau)x+g(\tau) \label{eq:elliptic_curve} 
\end{align}
where $f(\tau)$ and $g(\tau)$ is defined by the Eisenstein series as $f(\tau):=-\frac{\pi^4}{3}E_4(\tau)$ and $g(\tau):=-\frac{2\pi^6}{27}E_6(\tau)$.
Let $\Delta(\tau)$ denote the discriminant of the right-hand side of Eq.~\eqref{eq:elliptic_curve}:
\begin{align}
\Delta(\tau)=4f(\tau)^3+27g(\tau)^2.
\end{align}
When $\Delta(\tau)=0$, the elliptic curve is called singular at the locus.
In the F-theory framework, the discriminant loci 
denote the locations of the $7$-branes. 
Thus, if 
\begin{align}
\Delta\left(\tau\left(z_*\right)\right)=0,
\end{align}
then $7$-branes are placed on $z_*\in\Sigma$.
If $N$ $7$-branes are stacked at the location $z_*$, the total space of the fibration is also singular at the locus, and we have
\begin{align}
N
=
\underset{z\to z_*}{\rm ord}\left(\Delta\left(\tau\left(z\right)\right)\right).
    \label{eq:N=ord_delta}
\end{align}
Here, $\underset{z\to z_*}{\rm ord}(\cdots)$ denotes the order of vanishing of $\cdots$ at $z=z_*$.

We now construct the topological operator associated with the $8$-form symmetry $\bZ_{12}^{[8]}$ based on the framework.
In the previous section, we have seen that the charge objects of the $\U(1)^{[8]}$ symmetry are  D7-branes.
Motivated by this observation, we expect that charged objects of the $\bZ_{12}^{[8]}$ symmetry are $7$-branes.
Thus the topological operator we construct measures the number of the $7$-branes.
Let us introduce the following quantity:
\begin{align}
\mathcal{Q}(\gamma)=\frac{1}{2\pi {\rm i}}\oint_\gamma {\rm d}z\frac{{\rm d}}{{\rm d}z}\log\Delta(\tau(z)).
\label{eq:mathcal_Q}
\end{align}
If inside $\gamma$ there are $n$ points $z^{(1)},z^{(2)},\cdots,z^{(n)}$ at which $\Delta(\tau(z^{(i)}))=0$, then we have
\begin{align}
\mathcal{Q}(\gamma)
=
\sum_{i=1}^n\underset{z\to z^{(i)}}{\rm ord}\left(\Delta\left(\tau(z\right)\right)).
\end{align}
In other words, $\mathcal{Q}(\gamma)$ counts the number of 7-branes enclosed by $\gamma$.
Here we define the operator $\mathcal{W}_\phi(\gamma)$ as
\begin{align}\label{eq:mathcal_W}
\mathcal{W}_\phi(\gamma)
=
\exp({\rm i}\phi \mathcal{Q}(\gamma)).
\end{align}
By Cauchy's integral theorem, this is a topological operator.
However, the $\SL(2,\bZ)$ gauge invariance of $\mathcal{W}_\phi(\gamma)$ is not manifest.
In fact, as we see in the next paragraph, this operator turns out to be gauge invariant only if $\phi\in \frac{2\pi {\rm i}}{12}\bZ$.

Let us explicitly compute the $\SL(2,\bZ)$ gauge transformation of the operator.
Defining
\begin{align}\begin{aligned}
    \tau'(z)&:=\frac{a\tau(z)+b}{c\tau(z)+d},&
\begin{pmatrix}
   a & b \\
   c & d
\end{pmatrix}
&\in\SL(2,\bZ),
\end{aligned}\end{align}
we obtain
\begin{align}
\Delta(\tau'(z))
=
(c\tau(z)+d)^{12}\Delta(\tau(z)).
\end{align}
Note that  $\Delta(\tau)$ is a modular of weight $12$ since $E_4(\tau)$ and $E_6(\tau)$ have weight $4$ and $6$ respectively.
Thus, the gauge transformation of the operator $\mathcal{Q}(\gamma)$ is given by
\begin{align}
\mathcal{Q}'(\gamma)
=
\mathcal{Q}(\gamma)
+
12\times
\frac{1}{2\pi {\rm i}}\oint_\gamma {\rm d}z\frac{{\rm d}}{{\rm d}z}\log(c\tau(z)+d)
\end{align}
Regarding the second term on the right-hand side, in general, if there exists a point $z$ inside the region enclosed by $\gamma$ such that
\begin{align}
    \tau(z)=-\frac{d}{c},
\end{align}
then
\begin{align}
\frac{1}{2\pi {\rm i}}\oint_\gamma {\rm d}z\frac{{\rm d}}{{\rm d}z}\log(c\tau(z)+d)
\in \bZ.
\end{align}
Therefore, $\mathcal{Q}(\gamma)$ is well-defined modulo 12. We conclude
\begin{align}
   \phi= \frac{2\pi k}{12},
   \ 
   k\in\bZ/12\bZ,
\end{align}
and $\mathcal{W}_\phi(\gamma)$ generates the $\bZ_{12}$ $8$-form symmetry.

\section{Extensions of the duality group and emergent symmetries}
\label{sec:Extensions of the duality group and emergent symmetries}
In the previous section, we discuss the emergent symmetry based on the duality group $\SL(2,\bZ)$.
However, the true duality group is not $\SL(2,\bZ)$ but the extensions of it. 
In this section, we see the emergent symmetries, taking this into account.

\subsection{Extensions of the duality group}

In much of the literature, the duality group of type IIB superstring theory is described as $\SL(2,\bZ)$.
However, this is not precise, since there exist two possible extensions of the duality group~\cite{Dabholkar:1997zd,Pantev:2016nze,Tachikawa:2018njr,Debray:2021vob,Debray:2023yrs}.

One extension arises from the symmetry associated with the fermion number $F$ in the target space. The center of $\SL(2,\bZ)$ is 
\begin{align}
    \bZ_2=\{1,\mathscr{C}\},
\end{align}
and between this $\mathscr{C}$ and the fermion parity$(-1)^F$ there holds the relation,
\begin{align}
    \mathscr{C}^2=(-1)^F.
\end{align}
Thus, by taking into account the fermionic symmetry $(\bZ_2)_{F}$, one obtains a nontrivial extension:
\begin{align}\label{eq:Mp_extension}
0
\to
(\bZ_2)_{F}
\to
\Mp(2,\bZ)
\to
\SL(2,\bZ)
\to 
0.
\end{align}
In other words, the duality group is at least the double cover of $\SL(2,\bZ)$, namely $\Mp(2,\bZ)$.
Note that the nontrivial central extension of $\SL(2,\bZ)$ by $\bZ_{2}$ is unique, since $H^{2}(\SL(2,\bZ),\bZ_{2})\cong\bZ_{2}$.

The other extension arises from the symmetry associated with the worldsheet orientation reversal $\Omega$.
We denote this symmetry as $(\bZ_{2})_{\Omega}$, i.e.,
\begin{align}
(\bZ_2)_\Omega
=
\{1,\Omega\}.
\end{align}
For example, the orientation reversal $\Omega$ acts on the axio-dilaton as $\tau\mapsto -\bar{\tau}$.
This indicates that, since $(\bZ_2)_\Omega$ has a nontrivial action on $\SL(2,\bZ)$, they form the following semidirect product:\footnote{The worldsheet theory of type~IIB superstring also posseses the $(\bZ_{2})_{F_{L}}$ symmetry generated by the left-moving spacetime fermion number $(-1)^{F_L}$. This $(\bZ_{2})_{F_{L}}$ group is included in $\GL(2,\bZ)$ and inner isomorphic to $(\bZ_{2})_{\Omega}$.}
\begin{align}
\GL(2,\bZ)
=
\SL(2,\bZ)
\rtimes 
(\bZ_2)_\Omega.
\end{align}

Combining the above two extensions, one finds that the true duality group of type IIB superstring theory is the double cover of $\GL(2,\bZ)$.
\[
\begin{tikzcd}
\SL(2,\bZ)
\arrow[r,"\Omega"] \arrow[d,"(-1)^F"'] 
&
\GL(2,\bZ)
\arrow[d,""] \\
\Mp(2,\bZ) 
\arrow[r,""'] 
& 
\mbox{True duality\ group}
\end{tikzcd}
\]
Here, the square of $\Omega$ is not $(-1)^F$ but the identity,
\begin{align}
    \Omega^2=1.
\end{align}
Therefore, this double cover is of the $\Pin^+$ type rather than the $\Pin^-$ type.
Following~\cite{Debray:2021vob}, we refer to the true duality group as $\GL^{+}(2,\bZ)$:
\begin{align}
\mbox{True duality\ group}
=
\Pin^+(\GL(2,\bZ))
=
\GL^+(2,\bZ).
\end{align}

\subsection{Gauge or global symmetry}
\label{subsec:Gauge or global symmetry}

As stated in the outset of Section~\ref{sec:Quantum symmetry of SL(2,Z)}, superstring theory does not have any global symmetries.
The previous section, in short, concerns adding $(\bZ_2)_{F}$ and $(\bZ_2)_\Omega$ to $\SL(2,\bZ)$.
These $\bZ_2$ symmetries are also not global in superstring theory.

However, in this paper we would like to discuss the behavior of these symmetries not in superstring theory itself, but in its low-energy effective field theory, namely supergravity.
In typical backgrounds --- such as those considered in F-theory GUTs ---
the fermionic symmetry $(\bZ_2)_{F}$ appears as a gauge symmetry, 
while the worldsheet orientation reversal symmetry $(\bZ_2)_\Omega$ is often a global symmetry.

The reason why $(\bZ_2)_{F}$ is a gauge symmetry is that 7-branes provide a nontrivial $(\bZ_2)_{F}$ gauge field background.
As we will see later, a $\bZ_{24}^{[8]}$ global symmetry arises as the quantum symmetry of $\Mp(2,\bZ)$, and this is the proper solitonic symmetry associated with the 7-branes.

On the other hand, we consider backgrounds in which $(\bZ_2)_\Omega$ is realized as a global symmetry.
This is because, for example, $\Omega$ maps a D3-brane to an anti D3-brane.
This means that gauging $(\bZ_2)_\Omega$ allows configurations that break supersymmetry~\cite{Dierigl:2022reg,Debray:2023yrs}.
In this paper, we focus on the BPS configurations preserving the supersymmetry.
Thus, at low energies, we see $(\bZ_2^{[0]})_\Omega$ as an emergent 0-form global symmetry.

\subsection{Fermion parity and emergent symmetry}

Here, we consider the case where the $(\bZ_2)_{F}$ gauge symmetry is added to the $\SL(2,\bZ)$ gauge symmetry.
To consider the quantum symmetry of these, we apply Hom-functor $\Hom(-,\U(1))$ to the sequence~\eqref{eq:Mp_extension}:\footnote{
This functor is contravariant and left exact, but not right exact.
However, if we identify $\Hom(-,\U(1))\cong H^1(-,\U(1))$, then since $H^{i\ge 2}(-,\U(1))=0$, it can effectively be regarded as exact on both sides.
}
\begin{align}
0
\to
\Hom(\SL(2,\bZ),\U(1))
\to
\Hom(\Mp(2,\bZ),\U(1))
\to
\Hom((\bZ_2)_{F},\U(1))
\to 
0,
\end{align}
equivalently,
\begin{align}
0
\to
\bZ_{12}
\to
\bZ_{24}
\to
\bZ_{2}
\to 
0.
\end{align}
Therefore, in addition to the $\bZ_{12}^{[8]}$ global symmetry explained in Section \ref{subsec:Emergent symmetry associated with 7-brane}, there is also a $\bZ_{2}^{[8]}$ global symmetry, and together they form a $\bZ_{24}^{[8]}$ global symmetry.

The charged object of $\bZ_{24}^{[8]}$ is a 7-brane, since 12 7-branes give a nontrivial $(-1)^F$ monodromy.
Hence, the topological operator of $\bZ_{24}^{[8]}$ takes the following form:
\begin{align}
\mathscr{W}_\varphi(\gamma)
=
\exp({\rm i}\varphi\mathscr{Q}(\gamma)),
\end{align}
where
\begin{align}
   \varphi= \frac{2\pi k}{24},
   \ 
   k\in\bZ/24\bZ,
\end{align}
and $\mathscr{Q}(\gamma)$ is equal modulo 24 to the number of 7-branes inside $\gamma$.

Let us consider the selection rule originating from this 8-form global symmetry.
Suppose that $N$ 7-branes are aligned in parallel.
Let $\Sigma$ denote the surface transverse to the 7-branes, and assume that it is closed.
Then, in order for the expectation value to be nonvanishing, $N$ must be a multiple of 24.
This selection rule is natural, since each 7-brane contributes a deficit angle of $\pi/6$.
If $N=24$, then $\Sigma\cong \mathbb{CP}^1$, which is consistent with the selection rule.
The case that $N$ is larger than $24$ corresponds to the background where the supersymmetry is broken, which is discussed in \cite{Kleban:2007kk}.

\subsection{Worldsheet orientation reversal and emergent symmetry}

As mentioned in Section~\ref{subsec:Gauge or global symmetry}, 
we regard $(\bZ^{[0]}_2)_{\rm \Omega}$ as a global symmetry.
The emergent symmetries $(\bZ^{[0]}_2)_{\rm \Omega}$
and 
$\bZ_{12}^{[8]}$ (or $\bZ_{24}^{[8]}$)
are combined in a subtle way.

Let us denote the topological operator of $(\bZ^{[0]}_2)_{\rm \Omega}$ by $U(V)$,
where $V$ is a $(d-1)$-dimensional submanifold.
The operator $U(V)$ satisfies
\begin{align}
U(V)\cdot U(V)=1.
\end{align}

When $\gamma$ is sandwiched between $V$ and $V'$,
\begin{align}
U(V)
\mathcal{W}_\phi(\gamma)
U(V')
=
\mathcal{W}_{-\phi}(\gamma)
\end{align}
for the $\bZ_{12}^{[8]}$ symmetry
and 
\begin{align}
U(V)
\mathscr{W}_\varphi(\gamma)
U(V')
=
\mathscr{W}_{-\varphi}(\gamma)    
\end{align}
for the $\bZ_{24}^{[8]}$ symmetry. 
This is because $\Omega$ maps a 7-brane to an anti 7-brane.
Indeed, this can be checked explicitly for the expressions
\eqref{eq:mathcal_Q} and \eqref{eq:mathcal_W} with $\Omega:\tau\mapsto-\bar{\tau}$.

From the above, we find that type IIB supergravity possesses a global symmetry mixing a 0-form symmetry and an 8-form symmetry,\footnote{The structure of symmetry obtained by gauging finite normal abelian subgroup is studied in~\cite{Tachikawa:2017gyf}. This argument cannot be applied directly to the situation in this section, since the gauged symmetry $\SL(2,\bZ)$ is neither finite nor abelian. It would be interesting to extend the method developed in~\cite{Tachikawa:2017gyf} to include our current case.} namely 
$\bZ_{24}^{[8]}\rtimes (\bZ_2^{[0]})_\Omega$.
Such a structure, in which symmetries of different form degrees are intertwined, is often referred to as a higher-group symmetry~\cite{Green:1984sg,Kapustin:2013uxa,Cordova:2018cvg,Benini:2018reh}.\footnote{Higer-group global symmetries are extensively studied. See e.g.~\cite{Komargodski:2017dmc,Tanizaki:2019rbk,Hidaka:2019mfm,Hidaka:2020iaz,Hsin:2020nts,Cordova:2020tij,DelZotto:2020sop,Hidaka:2020izy,Bhardwaj:2021ojs,Apruzzi:2021vcu,Bhardwaj:2021wif,Hidaka:2021mml,Hidaka:2021kkf,Apruzzi:2021mlh,DelZotto:2022joo,Debray:2023rlx,Anber:2024gis}.}

\section{Discussions}
\label{sec:Discussions}

In this paper, we have considered type IIB supergravity with the global symmetry $\bZ^{[8]}_{12}$.
However, based on the discussion in Section \ref{sec:Brief review of quantum symmetry}, it turns out that there exist at least two variants.\footnote{
If taking the extensions of the duality group into account, there are more variants of type IIB supergravity.
}
One variant is the theory obtained by gauging $\bZ^{[8]}_{12}$.
This theory possesses a global symmetry $\bZ^{[0]}_{12}$ as its quantum symmetry.
Another variant is a theory in which the $\SL(2,\bZ)$ is not a gauge symmetry but a global symmetry.
As explained in Section \ref{subsec:General case}, in order to ungauge this symmetry, one needs to focus on the non-invertible symmetry ${\rm Rep}(\SL(2,\bZ))$.

To summarize, there exist several versions of type IIB supergravity with different global symmetries:
\begin{itemize}
\item[(1)] 
type IIB supergravity with
$\bZ_{12}^{[8]}$,
\item[(2)]
type IIB supergravity with
$\bZ_{12}^{[0]}$, and,
\item[(3)]
type IIB supergravity with
$\SL(2,\bZ)^{[0]}$.
\end{itemize}
The fact that the global symmetries differ means that these are not equivalent as quantum field theories. 
Indeed, all the massless degrees of freedom in these theories are the same. 
However, there exist subtle differences only in the heavy degrees of freedom (the topological phases).
In other words, taking the limit of the string length $\ell_{\rm string}\to 0$ does not fully specify the effective field theory.
This seems to reflect an essential distinction between string theory and quantum field theory.

In this paper, we have focused in particular on case (1), since we consider a theory consistent with F-theory.
Because this theory possesses nontrivial monodromies from the $\SL(2,\bZ)$ gauge symmetry, the F-theoretic description is available.
In other words, one may say that the conventional framework of F-theory GUTs implicitly involved a choice of the topological phase.

It may be interesting to investigate these subtle differences in topological phases in supergravity more deeply.\footnote{See e.g.~\cite{Garcia-Etxebarria:2017crf,Freed:2019sco,Debray:2021vob,Apruzzi:2021nmk,Lee:2022spd,Tachikawa:2024ucm,Braeger:2025kra,Tachikawa:2025flw,Chakrabhavi:2025bfi} for related studies. Particularly, in \cite{Debray:2021vob}, they claim that there are several kinds of type IIB theory in terms of the anomaly cancellation. 
These type IIB theories have different massive spectra and extended objects. 
This argument seems to be similar to our discussion.
We leave this point for the future work.}
Such a study may, for example, reveal differences in the types of branes contained in the theory.
To this end, it is necessary to carefully analyze the non-invertible symmetry ${\rm Rep}(\SL(2,\bZ))$.

Moreover, the discussions in this paper mainly depend only on symmetry. Therefore, it can be applied to other theories with similar symmetries.

Type 0B string theory\cite{Dixon:1986iz} is one such example.
In \cite{Bergman:1999km}, it has been conjectured that this theory has a connection with M-theory.
If one accepts this conjecture, then, since type 0B theory possesses an $\SL(2,\bZ)$ symmetry, it is expected to have a $\bZ_{12}^{[8]} $ symmetry just as type IIB theory does.

As another possible future direction, this research might be applicable to string phenomenology. The reason is that global symmetries or topological operators are invariant under the renormalization group flow~\cite{Gaiotto:2014kfa,Gaiotto:2017yup}. Therefore, the $\bZ_{24}^{[8]}\rtimes (\bZ_{2}^{[0]})_\Omega$ symmetry must also exist in the low-energy theory.
Since this structure is independent of the details of compactification, it may provide a model-independent constraint.

\section*{Acknowledgements}
The authors would like to thank Y.~Hamada, T.~Kimura, Y. ~Moriwaki, Y.~Nakayama, S.~Sasaki, S.~Shimamori, S.~Sugimoto, S.~Yamaguchi, and R.~Yokokura for helpful discussions.
The authors are especially grateful to Y.~Tanizaki and K.~Yonekura for valuable comments.
The work of NK is supported by JSPS KAKENHI Grant Number JP24KJ0157.
The work of MK is supported by JSPS KAKENHI Grant Number JP23K22491, 25K23381.
The work of HW is supported by JSPS KAKENHI Grant-in-Aid for JSPS fellows Grant Number 24KJ1603.

\bibliographystyle{ytphys}
%\baselineskip=.95\baselineskip
\bibliography{main}

\providecommand{\href}[2]{#2}\begingroup\raggedright\begin{thebibliography}{10}

\bibitem{Scherk:1974ca}
J.~Scherk and J.~H. Schwarz, {\slshape {Dual Models for Nonhadrons},} \href{http://dx.doi.org/10.1016/0550-3213(74)90010-8}{{\em Nucl. Phys. B} {\bfseries 81} (1974) 118--144}.

\bibitem{Veneziano:1968yb}
G.~Veneziano, {\slshape {Construction of a crossing - symmetric, Regge behaved amplitude for linearly rising trajectories},} \href{http://dx.doi.org/10.1007/BF02824451}{{\em Nuovo Cim. A} {\bfseries 57} (1968) 190--197}.

\bibitem{Banks:1988yz}
T.~Banks and L.~J. Dixon, {\slshape {Constraints on String Vacua with Space-Time Supersymmetry},} \href{http://dx.doi.org/10.1016/0550-3213(88)90523-8}{{\em Nucl. Phys. B} {\bfseries 307} (1988) 93--108}.

\bibitem{Banks:2010zn}
T.~Banks and N.~Seiberg, {\slshape {Symmetries and Strings in Field Theory and Gravity},} \href{http://dx.doi.org/10.1103/PhysRevD.83.084019}{{\em Phys. Rev. D} {\bfseries 83} (2011) 084019}, \href{http://arxiv.org/abs/1011.5120}{{ arXiv:1011.5120~[hep-th]}}.

\bibitem{tHooft:1979rat}
G.~'t~Hooft, {\slshape {Naturalness, chiral symmetry, and spontaneous chiral symmetry breaking},} \href{http://dx.doi.org/10.1007/978-1-4684-7571-5_9}{{\em NATO Sci. Ser. B} {\bfseries 59} (1980) 135--157}.

\bibitem{Vafa:1996xn}
C.~Vafa, {\slshape {Evidence for F theory},} \href{http://dx.doi.org/10.1016/0550-3213(96)00172-1}{{\em Nucl. Phys. B} {\bfseries 469} (1996) 403--418}, \href{http://arxiv.org/abs/hep-th/9602022}{{ arXiv:hep-th/9602022}}.

\bibitem{Donagi:2008ca}
R.~Donagi and M.~Wijnholt, {\slshape {Model Building with F-Theory},} \href{http://dx.doi.org/10.4310/ATMP.2011.v15.n5.a2}{{\em Adv. Theor. Math. Phys.} {\bfseries 15} (2011) 1237--1317}, \href{http://arxiv.org/abs/0802.2969}{{ arXiv:0802.2969~[hep-th]}}.

\bibitem{Beasley:2008dc}
C.~Beasley, J.~J. Heckman, and C.~Vafa, {\slshape {GUTs and Exceptional Branes in F-theory - I},} \href{http://dx.doi.org/10.1088/1126-6708/2009/01/058}{{\em JHEP} {\bfseries 01} (2009) 058}, \href{http://arxiv.org/abs/0802.3391}{{ arXiv:0802.3391~[hep-th]}}.

\bibitem{Beasley:2008kw}
C.~Beasley, J.~J. Heckman, and C.~Vafa, {\slshape {GUTs and Exceptional Branes in F-theory - II: Experimental Predictions},} \href{http://dx.doi.org/10.1088/1126-6708/2009/01/059}{{\em JHEP} {\bfseries 01} (2009) 059}, \href{http://arxiv.org/abs/0806.0102}{{ arXiv:0806.0102~[hep-th]}}.

\bibitem{Donagi:2008kj}
R.~Donagi and M.~Wijnholt, {\slshape {Breaking GUT Groups in F-Theory},} \href{http://dx.doi.org/10.4310/ATMP.2011.v15.n6.a1}{{\em Adv. Theor. Math. Phys.} {\bfseries 15} (2011) 1523--1603}, \href{http://arxiv.org/abs/0808.2223}{{ arXiv:0808.2223~[hep-th]}}.

\bibitem{Kapustin:2014gua}
A.~Kapustin and N.~Seiberg, {\slshape {Coupling a QFT to a TQFT and Duality},} \href{http://dx.doi.org/10.1007/JHEP04(2014)001}{{\em JHEP} {\bfseries 04} (2014) 001}, \href{http://arxiv.org/abs/1401.0740}{{ arXiv:1401.0740~[hep-th]}}.

\bibitem{Gaiotto:2014kfa}
D.~Gaiotto, A.~Kapustin, N.~Seiberg, and B.~Willett, {\slshape {Generalized Global Symmetries},} \href{http://dx.doi.org/10.1007/JHEP02(2015)172}{{\em JHEP} {\bfseries 02} (2015) 172}, \href{http://arxiv.org/abs/1412.5148}{{ arXiv:1412.5148~[hep-th]}}.

\bibitem{DelZotto:2015isa}
M.~Del~Zotto, J.~J. Heckman, D.~S. Park, and T.~Rudelius, {\slshape {On the Defect Group of a 6D SCFT},} \href{http://dx.doi.org/10.1007/s11005-016-0839-5}{{\em Lett. Math. Phys.} {\bfseries 106} (2016) 765--786}, \href{http://arxiv.org/abs/1503.04806}{{ arXiv:1503.04806~[hep-th]}}.

\bibitem{Morrison:2020ool}
D.~R. Morrison, S.~Schafer-Nameki, and B.~Willett, {\slshape {Higher-Form Symmetries in 5d},} \href{http://dx.doi.org/10.1007/JHEP09(2020)024}{{\em JHEP} {\bfseries 09} (2020) 024}, \href{http://arxiv.org/abs/2005.12296}{{ arXiv:2005.12296~[hep-th]}}.

\bibitem{Albertini:2020mdx}
F.~Albertini, M.~Del~Zotto, I.~Garc{\'\i}a~Etxebarria, and S.~S. Hosseini, {\slshape {Higher Form Symmetries and M-theory},} \href{http://dx.doi.org/10.1007/JHEP12(2020)203}{{\em JHEP} {\bfseries 12} (2020) 203}, \href{http://arxiv.org/abs/2005.12831}{{ arXiv:2005.12831~[hep-th]}}.

\bibitem{DelZotto:2020esg}
M.~Del~Zotto, I.~Garc{\'\i}a~Etxebarria, and S.~S. Hosseini, {\slshape {Higher form symmetries of Argyres-Douglas theories},} \href{http://dx.doi.org/10.1007/JHEP10(2020)056}{{\em JHEP} {\bfseries 10} (2020) 056}, \href{http://arxiv.org/abs/2007.15603}{{ arXiv:2007.15603~[hep-th]}}.

\bibitem{Bhardwaj:2020phs}
L.~Bhardwaj and S.~Sch{\"a}fer-Nameki, {\slshape {Higher-form symmetries of 6d and 5d theories},} \href{http://dx.doi.org/10.1007/JHEP02(2021)159}{{\em JHEP} {\bfseries 02} (2021) 159}, \href{http://arxiv.org/abs/2008.09600}{{ arXiv:2008.09600~[hep-th]}}.

\bibitem{Gukov:2020btk}
S.~Gukov, P.-S. Hsin, and D.~Pei, {\slshape {Generalized global symmetries of $T[M]$ theories. Part I},} \href{http://dx.doi.org/10.1007/JHEP04(2021)232}{{\em JHEP} {\bfseries 04} (2021) 232}, \href{http://arxiv.org/abs/2010.15890}{{ arXiv:2010.15890~[hep-th]}}.

\bibitem{Heidenreich:2020pkc}
B.~Heidenreich, J.~McNamara, M.~Montero, M.~Reece, T.~Rudelius, and I.~Valenzuela, {\slshape {Chern-Weil global symmetries and how quantum gravity avoids them},} \href{http://dx.doi.org/10.1007/JHEP11(2021)053}{{\em JHEP} {\bfseries 11} (2021) 053}, \href{http://arxiv.org/abs/2012.00009}{{ arXiv:2012.00009~[hep-th]}}.

\bibitem{Apruzzi:2021nmk}
F.~Apruzzi, F.~Bonetti, I.~Garc{\'\i}a~Etxebarria, S.~S. Hosseini, and S.~Schafer-Nameki, {\slshape {Symmetry TFTs from String Theory},} \href{http://dx.doi.org/10.1007/s00220-023-04737-2}{{\em Commun. Math. Phys.} {\bfseries 402} (2023) 895--949}, \href{http://arxiv.org/abs/2112.02092}{{ arXiv:2112.02092~[hep-th]}}.

\bibitem{Bhardwaj:2021mzl}
L.~Bhardwaj, S.~Giacomelli, M.~H{\"u}bner, and S.~Sch{\"a}fer-Nameki, {\slshape {Relative defects in relative theories: Trapped higher-form symmetries and irregular punctures in class S},} \href{http://dx.doi.org/10.21468/SciPostPhys.13.4.101}{{\em SciPost Phys.} {\bfseries 13} (2022) 101}, \href{http://arxiv.org/abs/2201.00018}{{ arXiv:2201.00018~[hep-th]}}.

\bibitem{Genolini:2022mpi}
P.~B. Genolini and L.~Tizzano, {\slshape {Comments on Global Symmetries and Anomalies of 5d SCFTs},} \href{http://dx.doi.org/10.1007/s00220-024-05139-8}{{\em Commun. Math. Phys.} {\bfseries 405} (2024) 255}, \href{http://arxiv.org/abs/2201.02190}{{ arXiv:2201.02190~[hep-th]}}.

\bibitem{DelZotto:2022fnw}
M.~Del~Zotto, J.~J. Heckman, S.~N. Meynet, R.~Moscrop, and H.~Y. Zhang, {\slshape {Higher symmetries of 5D orbifold SCFTs},} \href{http://dx.doi.org/10.1103/PhysRevD.106.046010}{{\em Phys. Rev. D} {\bfseries 106} (2022) 046010}, \href{http://arxiv.org/abs/2201.08372}{{ arXiv:2201.08372~[hep-th]}}.

\bibitem{Cvetic:2022imb}
M.~Cveti{\v{c}}, J.~J. Heckman, M.~H{\"u}bner, and E.~Torres, {\slshape {0-form, 1-form, and 2-group symmetries via cutting and gluing of orbifolds},} \href{http://dx.doi.org/10.1103/PhysRevD.106.106003}{{\em Phys. Rev. D} {\bfseries 106} (2022) 106003}, \href{http://arxiv.org/abs/2203.10102}{{ arXiv:2203.10102~[hep-th]}}.

\bibitem{Apruzzi:2022dlm}
F.~Apruzzi, {\slshape {Higher form symmetries TFT in 6d},} \href{http://dx.doi.org/10.1007/JHEP11(2022)050}{{\em JHEP} {\bfseries 11} (2022) 050}, \href{http://arxiv.org/abs/2203.10063}{{ arXiv:2203.10063~[hep-th]}}.

\bibitem{Hubner:2022kxr}
M.~Hubner, D.~R. Morrison, S.~Schafer-Nameki, and Y.-N. Wang, {\slshape {Generalized Symmetries in F-theory and the Topology of Elliptic Fibrations},} \href{http://dx.doi.org/10.21468/SciPostPhys.13.2.030}{{\em SciPost Phys.} {\bfseries 13} (2022) 030}, \href{http://arxiv.org/abs/2203.10022}{{ arXiv:2203.10022~[hep-th]}}.

\bibitem{Heckman:2022muc}
J.~J. Heckman, M.~H{\"u}bner, E.~Torres, and H.~Y. Zhang, {\slshape {The Branes Behind Generalized Symmetry Operators},} \href{http://dx.doi.org/10.1002/prop.202200180}{{\em Fortsch. Phys.} {\bfseries 71} (2023) 2200180}, \href{http://arxiv.org/abs/2209.03343}{{ arXiv:2209.03343~[hep-th]}}.

\bibitem{Grimm:2022xmj}
T.~W. Grimm, S.~Lanza, and T.~van Vuren, {\slshape {Global symmetry-breaking and generalized theta-terms in Type IIB EFTs},} \href{http://dx.doi.org/10.1007/JHEP10(2023)154}{{\em JHEP} {\bfseries 10} (2023) 154}, \href{http://arxiv.org/abs/2211.11769}{{ arXiv:2211.11769~[hep-th]}}.

\bibitem{Etheredge:2023ler}
M.~Etheredge, I.~Garcia~Etxebarria, B.~Heidenreich, and S.~Rauch, {\slshape {Branes and symmetries for $ \mathcal{N} $ = 3 S-folds},} \href{http://dx.doi.org/10.1007/JHEP09(2023)005}{{\em JHEP} {\bfseries 09} (2023) 005}, \href{http://arxiv.org/abs/2302.14068}{{ arXiv:2302.14068~[hep-th]}}.

\bibitem{Amariti:2023hev}
A.~Amariti, D.~Morgante, A.~Pasternak, S.~Rota, and V.~Tatitscheff, {\slshape {One-form symmetries in $\mathcal{N} = 3$ S-folds},} \href{http://dx.doi.org/10.21468/SciPostPhys.15.4.132}{{\em SciPost Phys.} {\bfseries 15} (2023) 132}, \href{http://arxiv.org/abs/2303.07299}{{ arXiv:2303.07299~[hep-th]}}.

\bibitem{Cvetic:2023pgm}
M.~Cveti{\v{c}}, J.~J. Heckman, M.~H{\"u}bner, and E.~Torres, {\slshape {Generalized symmetries, gravity, and the swampland},} \href{http://dx.doi.org/10.1103/PhysRevD.109.026012}{{\em Phys. Rev. D} {\bfseries 109} (2024) 026012}, \href{http://arxiv.org/abs/2307.13027}{{ arXiv:2307.13027~[hep-th]}}.

\bibitem{Lawrie:2023uiu}
C.~Lawrie and L.~Mansi, {\slshape {Higgs branch of heterotic little string theories: Hasse diagrams and generalized symmetries},} \href{http://dx.doi.org/10.1103/PhysRevD.110.026016}{{\em Phys. Rev. D} {\bfseries 110} (2024) 026016}, \href{http://arxiv.org/abs/2312.05306}{{ arXiv:2312.05306~[hep-th]}}.

\bibitem{DelZotto:2024tae}
M.~Del~Zotto, S.~N. Meynet, and R.~Moscrop, {\slshape {Remarks on geometric engineering, symmetry TFTs and anomalies},} \href{http://dx.doi.org/10.1007/JHEP07(2024)220}{{\em JHEP} {\bfseries 07} (2024) 220}, \href{http://arxiv.org/abs/2402.18646}{{ arXiv:2402.18646~[hep-th]}}.

\bibitem{Braeger:2024jcj}
N.~Braeger, V.~Chakrabhavi, J.~J. Heckman, and M.~H{\"u}bner, {\slshape {Generalized symmetries of nonsupersymmetric orbifolds},} \href{http://dx.doi.org/10.1103/PhysRevD.111.066015}{{\em Phys. Rev. D} {\bfseries 111} (2025) 066015}, \href{http://arxiv.org/abs/2404.17639}{{ arXiv:2404.17639~[hep-th]}}.

\bibitem{Franco:2024mxa}
S.~Franco and X.~Yu, {\slshape {Generalized symmetries in 2D from string theory: SymTFTs, intrinsic relativeness, and anomalies of non-invertible symmetries},} \href{http://dx.doi.org/10.1007/JHEP11(2024)004}{{\em JHEP} {\bfseries 11} (2024) 004}, \href{http://arxiv.org/abs/2404.19761}{{ arXiv:2404.19761~[hep-th]}}.

\bibitem{Baume:2024oqn}
F.~Baume, P.-K. Oehlmann, and F.~Ruehle, {\slshape {Bounds and dualities of Type II Little String Theories},} \href{http://dx.doi.org/10.1007/JHEP11(2024)149}{{\em JHEP} {\bfseries 11} (2024) 149}, \href{http://arxiv.org/abs/2405.03877}{{ arXiv:2405.03877~[hep-th]}}.

\bibitem{Cvetic:2024mtt}
M.~Cveti{\v{c}}, M.~Dierigl, L.~Lin, E.~Torres, and H.~Y. Zhang, {\slshape {Frozen generalized symmetries},} \href{http://dx.doi.org/10.1103/PhysRevD.111.026018}{{\em Phys. Rev. D} {\bfseries 111} (2025) 026018}, \href{http://arxiv.org/abs/2410.07318}{{ arXiv:2410.07318~[hep-th]}}.

\bibitem{Tian:2024dgl}
J.~Tian and Y.-N. Wang, {\slshape {A tale of bulk and branes: Symmetry TFT of 6D SCFTs from IIB/F-theory},} \href{http://dx.doi.org/10.1007/JHEP03(2025)085}{{\em JHEP} {\bfseries 03} (2025) 085}, \href{http://arxiv.org/abs/2410.23076}{{ arXiv:2410.23076~[hep-th]}}.

\bibitem{Braeger:2025rov}
N.~Braeger, V.~Chakrabhavi, J.~J. Heckman, and M.~H{\"u}bner, {\slshape {Generalized Symmetries of Non-SUSY and Discrete Torsion String Backgrounds},} \href{http://arxiv.org/abs/2504.10484}{{ arXiv:2504.10484~[hep-th]}}.

\bibitem{Najjar:2024vmm}
M.~Najjar, L.~Santilli, and Y.-N. Wang, {\slshape {({\ensuremath{-}}1)-form symmetries from M-theory and SymTFTs},} \href{http://dx.doi.org/10.1007/JHEP03(2025)134}{{\em JHEP} {\bfseries 03} (2025) 134}, \href{http://arxiv.org/abs/2411.19683}{{ arXiv:2411.19683~[hep-th]}}.

\bibitem{Najjar:2025htp}
M.~Najjar, {\slshape {Modified instanton sum and 4-group structure in 4d $\mathcal{N}=1$$SU(M)$ SYM from holography},} \href{http://arxiv.org/abs/2503.17108}{{ arXiv:2503.17108~[hep-th]}}.

\bibitem{Khlaif:2025jnx}
O.~Khlaif and M.~Najjar, {\slshape {Aspects of 4d $\mathcal{N}=1$$ADE$ gauge theories from M-theory: decomposition, automorphisms, and generalised symmetries},} \href{http://arxiv.org/abs/2508.00564}{{ arXiv:2508.00564~[hep-th]}}.

\bibitem{Vafa:1989ih}
C.~Vafa, {\slshape {Quantum Symmetries of String Vacua},} \href{http://dx.doi.org/10.1142/S0217732389001842}{{\em Mod. Phys. Lett. A} {\bfseries 4} (1989) 1615}.

\bibitem{Dierigl:2020lai}
M.~Dierigl and J.~J. Heckman, {\slshape {Swampland cobordism conjecture and non-Abelian duality groups},} \href{http://dx.doi.org/10.1103/PhysRevD.103.066006}{{\em Phys. Rev. D} {\bfseries 103} (2021) 066006}, \href{http://arxiv.org/abs/2012.00013}{{ arXiv:2012.00013~[hep-th]}}.

\bibitem{Bhardwaj:2017xup}
L.~Bhardwaj and Y.~Tachikawa, {\slshape {On finite symmetries and their gauging in two dimensions},} \href{http://dx.doi.org/10.1007/JHEP03(2018)189}{{\em JHEP} {\bfseries 03} (2018) 189}, \href{http://arxiv.org/abs/1704.02330}{{ arXiv:1704.02330~[hep-th]}}.

\bibitem{Green:1982tk}
M.~B. Green and J.~H. Schwarz, {\slshape {Extended Supergravity in Ten-Dimensions},} \href{http://dx.doi.org/10.1016/0370-2693(83)90781-5}{{\em Phys. Lett. B} {\bfseries 122} (1983) 143--147}.

\bibitem{Hull:1994ys}
C.~M. Hull and P.~K. Townsend, {\slshape {Unity of superstring dualities},} \href{http://dx.doi.org/10.1201/9781482268737-24}{{\em Nucl. Phys. B} {\bfseries 438} (1995) 109--137}, \href{http://arxiv.org/abs/hep-th/9410167}{{ arXiv:hep-th/9410167}}.

\bibitem{Debray:2021vob}
A.~Debray, M.~Dierigl, J.~J. Heckman, and M.~Montero, {\slshape {The anomaly that was not meant IIB},} \href{http://dx.doi.org/10.1002/prop.202100168}{{\em Fortsch. Phys.} {\bfseries 70} (2022) 2100168}, \href{http://arxiv.org/abs/2107.14227}{{ arXiv:2107.14227~[hep-th]}}.

\bibitem{Mermin:1979zz}
N.~D. Mermin, {\slshape {The topological theory of defects in ordered media},} \href{http://dx.doi.org/10.1103/RevModPhys.51.591}{{\em Rev. Mod. Phys.} {\bfseries 51} (1979) 591--648}.

\bibitem{Chen:2022cyw}
S.~Chen and Y.~Tanizaki, {\slshape {Solitonic Symmetry beyond Homotopy: Invertibility from Bordism and Noninvertibility from Topological Quantum Field Theory},} \href{http://dx.doi.org/10.1103/PhysRevLett.131.011602}{{\em Phys. Rev. Lett.} {\bfseries 131} (2023) 011602}, \href{http://arxiv.org/abs/2210.13780}{{ arXiv:2210.13780~[hep-th]}}.

\bibitem{Seiberg:2018ntt}
N.~Seiberg, Y.~Tachikawa, and K.~Yonekura, {\slshape {Anomalies of Duality Groups and Extended Conformal Manifolds},} \href{http://dx.doi.org/10.1093/ptep/pty069}{{\em PTEP} {\bfseries 2018} (2018) 073B04}, \href{http://arxiv.org/abs/1803.07366}{{ arXiv:1803.07366~[hep-th]}}.

\bibitem{McNamara:2021cuo}
J.~McNamara, {\slshape {Gravitational Solitons and Completeness},} \href{http://arxiv.org/abs/2108.02228}{{ arXiv:2108.02228~[hep-th]}}.

\bibitem{Taylor:2011wt}
W.~Taylor, {\slshape {TASI Lectures on Supergravity and String Vacua in Various Dimensions},} \href{http://arxiv.org/abs/1104.2051}{{ arXiv:1104.2051~[hep-th]}}.

\bibitem{Weigand:2018rez}
T.~Weigand, {\slshape {F-theory},} {\em PoS} {\bfseries TASI2017} (2018) 016, \href{http://arxiv.org/abs/1806.01854}{{ arXiv:1806.01854~[hep-th]}}.

\bibitem{Dabholkar:1997zd}
A.~Dabholkar, {\slshape {Lectures on orientifolds and duality},} in {\em {ICTP Summer School in High-Energy Physics and Cosmology}}, pp.~128--191.
\newblock 6, 1997.
\newblock \href{http://arxiv.org/abs/hep-th/9804208}{{ arXiv:hep-th/9804208}}.

\bibitem{Pantev:2016nze}
T.~Pantev and E.~Sharpe, {\slshape {Duality group actions on fermions},} \href{http://dx.doi.org/10.1007/JHEP11(2016)171}{{\em JHEP} {\bfseries 11} (2016) 171}, \href{http://arxiv.org/abs/1609.00011}{{ arXiv:1609.00011~[hep-th]}}.

\bibitem{Tachikawa:2018njr}
Y.~Tachikawa and K.~Yonekura, {\slshape {Why are fractional charges of orientifolds compatible with Dirac quantization?},} \href{http://dx.doi.org/10.21468/SciPostPhys.7.5.058}{{\em SciPost Phys.} {\bfseries 7} (2019) 058}, \href{http://arxiv.org/abs/1805.02772}{{ arXiv:1805.02772~[hep-th]}}.

\bibitem{Debray:2023yrs}
A.~Debray, M.~Dierigl, J.~J. Heckman, and M.~Montero, {\slshape {The Chronicles of IIBordia: Dualities, Bordisms, and the Swampland},} \href{http://arxiv.org/abs/2302.00007}{{ arXiv:2302.00007~[hep-th]}}.

\bibitem{Dierigl:2022reg}
M.~Dierigl, J.~J. Heckman, M.~Montero, and E.~Torres, {\slshape {IIB string theory explored: Reflection 7-branes},} \href{http://dx.doi.org/10.1103/PhysRevD.107.086015}{{\em Phys. Rev. D} {\bfseries 107} (2023) 086015}, \href{http://arxiv.org/abs/2212.05077}{{ arXiv:2212.05077~[hep-th]}}.

\bibitem{Kleban:2007kk}
M.~Kleban and M.~Redi, {\slshape {Expanding F-Theory},} \href{http://dx.doi.org/10.1088/1126-6708/2007/09/038}{{\em JHEP} {\bfseries 09} (2007) 038}, \href{http://arxiv.org/abs/0705.2020}{{ arXiv:0705.2020~[hep-th]}}.

\bibitem{Tachikawa:2017gyf}
Y.~Tachikawa, {\slshape {On gauging finite subgroups},} \href{http://dx.doi.org/10.21468/SciPostPhys.8.1.015}{{\em SciPost Phys.} {\bfseries 8} (2020) 015}, \href{http://arxiv.org/abs/1712.09542}{{ arXiv:1712.09542~[hep-th]}}.

\bibitem{Green:1984sg}
M.~B. Green and J.~H. Schwarz, {\slshape {Anomaly Cancellation in Supersymmetric D=10 Gauge Theory and Superstring Theory},} \href{http://dx.doi.org/10.1016/0370-2693(84)91565-X}{{\em Phys. Lett. B} {\bfseries 149} (1984) 117--122}.

\bibitem{Kapustin:2013uxa}
A.~Kapustin and R.~Thorngren, {\slshape {Higher Symmetry and Gapped Phases of Gauge Theories},} \href{http://dx.doi.org/10.1007/978-3-319-59939-7_5}{{\em Prog. Math.} {\bfseries 324} (2017) 177--202}, \href{http://arxiv.org/abs/1309.4721}{{ arXiv:1309.4721~[hep-th]}}.

\bibitem{Cordova:2018cvg}
C.~C{\'o}rdova, T.~T. Dumitrescu, and K.~Intriligator, {\slshape {Exploring 2-Group Global Symmetries},} \href{http://dx.doi.org/10.1007/JHEP02(2019)184}{{\em JHEP} {\bfseries 02} (2019) 184}, \href{http://arxiv.org/abs/1802.04790}{{ arXiv:1802.04790~[hep-th]}}.

\bibitem{Benini:2018reh}
F.~Benini, C.~C{\'o}rdova, and P.-S. Hsin, {\slshape {On 2-Group Global Symmetries and their Anomalies},} \href{http://dx.doi.org/10.1007/JHEP03(2019)118}{{\em JHEP} {\bfseries 03} (2019) 118}, \href{http://arxiv.org/abs/1803.09336}{{ arXiv:1803.09336~[hep-th]}}.

\bibitem{Komargodski:2017dmc}
Z.~Komargodski, A.~Sharon, R.~Thorngren, and X.~Zhou, {\slshape {Comments on Abelian Higgs Models and Persistent Order},} \href{http://dx.doi.org/10.21468/SciPostPhys.6.1.003}{{\em SciPost Phys.} {\bfseries 6} (2019) 003}, \href{http://arxiv.org/abs/1705.04786}{{ arXiv:1705.04786~[hep-th]}}.

\bibitem{Tanizaki:2019rbk}
Y.~Tanizaki and M.~{\"U}nsal, {\slshape {Modified instanton sum in QCD and higher-groups},} \href{http://dx.doi.org/10.1007/JHEP03(2020)123}{{\em JHEP} {\bfseries 03} (2020) 123}, \href{http://arxiv.org/abs/1912.01033}{{ arXiv:1912.01033~[hep-th]}}.

\bibitem{Hidaka:2019mfm}
Y.~Hidaka, M.~Nitta, and R.~Yokokura, {\slshape {Emergent discrete 3-form symmetry and domain walls},} \href{http://dx.doi.org/10.1016/j.physletb.2020.135290}{{\em Phys. Lett. B} {\bfseries 803} (2020) 135290}, \href{http://arxiv.org/abs/1912.02782}{{ arXiv:1912.02782~[hep-th]}}.

\bibitem{Hidaka:2020iaz}
Y.~Hidaka, M.~Nitta, and R.~Yokokura, {\slshape {Higher-form symmetries and 3-group in axion electrodynamics},} \href{http://dx.doi.org/10.1016/j.physletb.2020.135672}{{\em Phys. Lett. B} {\bfseries 808} (2020) 135672}, \href{http://arxiv.org/abs/2006.12532}{{ arXiv:2006.12532~[hep-th]}}.

\bibitem{Hsin:2020nts}
P.-S. Hsin and H.~T. Lam, {\slshape {Discrete theta angles, symmetries and anomalies},} \href{http://dx.doi.org/10.21468/SciPostPhys.10.2.032}{{\em SciPost Phys.} {\bfseries 10} (2021) 032}, \href{http://arxiv.org/abs/2007.05915}{{ arXiv:2007.05915~[hep-th]}}.

\bibitem{Cordova:2020tij}
C.~Cordova, T.~T. Dumitrescu, and K.~Intriligator, {\slshape {2-Group Global Symmetries and Anomalies in Six-Dimensional Quantum Field Theories},} \href{http://dx.doi.org/10.1007/JHEP04(2021)252}{{\em JHEP} {\bfseries 04} (2021) 252}, \href{http://arxiv.org/abs/2009.00138}{{ arXiv:2009.00138~[hep-th]}}.

\bibitem{DelZotto:2020sop}
M.~Del~Zotto and K.~Ohmori, {\slshape {2-Group Symmetries of 6D Little String Theories and T-Duality},} \href{http://dx.doi.org/10.1007/s00023-021-01018-3}{{\em Annales Henri Poincare} {\bfseries 22} (2021) 2451--2474}, \href{http://arxiv.org/abs/2009.03489}{{ arXiv:2009.03489~[hep-th]}}.

\bibitem{Hidaka:2020izy}
Y.~Hidaka, M.~Nitta, and R.~Yokokura, {\slshape {Global 3-group symmetry and 't Hooft anomalies in axion electrodynamics},} \href{http://dx.doi.org/10.1007/JHEP01(2021)173}{{\em JHEP} {\bfseries 01} (2021) 173}, \href{http://arxiv.org/abs/2009.14368}{{ arXiv:2009.14368~[hep-th]}}.

\bibitem{Bhardwaj:2021ojs}
L.~Bhardwaj, {\slshape {Global form of flavor symmetry groups in 4d N=2 theories of class S},} \href{http://dx.doi.org/10.21468/SciPostPhys.12.6.183}{{\em SciPost Phys.} {\bfseries 12} (2022) 183}, \href{http://arxiv.org/abs/2105.08730}{{ arXiv:2105.08730~[hep-th]}}.

\bibitem{Apruzzi:2021vcu}
F.~Apruzzi, S.~Schafer-Nameki, L.~Bhardwaj, and J.~Oh, {\slshape {The Global Form of Flavor Symmetries and 2-Group Symmetries in 5d SCFTs},} \href{http://dx.doi.org/10.21468/SciPostPhys.13.2.024}{{\em SciPost Phys.} {\bfseries 13} (2022) 024}, \href{http://arxiv.org/abs/2105.08724}{{ arXiv:2105.08724~[hep-th]}}.

\bibitem{Bhardwaj:2021wif}
L.~Bhardwaj, {\slshape {2-Group symmetries in class S},} \href{http://dx.doi.org/10.21468/SciPostPhys.12.5.152}{{\em SciPost Phys.} {\bfseries 12} (2022) 152}, \href{http://arxiv.org/abs/2107.06816}{{ arXiv:2107.06816~[hep-th]}}.

\bibitem{Hidaka:2021mml}
Y.~Hidaka, M.~Nitta, and R.~Yokokura, {\slshape {Topological axion electrodynamics and 4-group symmetry},} \href{http://dx.doi.org/10.1016/j.physletb.2021.136762}{{\em Phys. Lett. B} {\bfseries 823} (2021) 136762}, \href{http://arxiv.org/abs/2107.08753}{{ arXiv:2107.08753~[hep-th]}}.

\bibitem{Hidaka:2021kkf}
Y.~Hidaka, M.~Nitta, and R.~Yokokura, {\slshape {Global 4-group symmetry and {\textquoteright}t Hooft anomalies in topological axion electrodynamics},} \href{http://dx.doi.org/10.1093/ptep/ptab150}{{\em PTEP} {\bfseries 2022} (2022) 04A109}, \href{http://arxiv.org/abs/2108.12564}{{ arXiv:2108.12564~[hep-th]}}.

\bibitem{Apruzzi:2021mlh}
F.~Apruzzi, L.~Bhardwaj, D.~S.~W. Gould, and S.~Schafer-Nameki, {\slshape {2-Group symmetries and their classification in 6d},} \href{http://dx.doi.org/10.21468/SciPostPhys.12.3.098}{{\em SciPost Phys.} {\bfseries 12} (2022) 098}, \href{http://arxiv.org/abs/2110.14647}{{ arXiv:2110.14647~[hep-th]}}.

\bibitem{DelZotto:2022joo}
M.~Del~Zotto, I.~Garc{\'\i}a~Etxebarria, and S.~Schafer-Nameki, {\slshape {2-Group Symmetries and M-Theory},} \href{http://dx.doi.org/10.21468/SciPostPhys.13.5.105}{{\em SciPost Phys.} {\bfseries 13} (2022) 105}, \href{http://arxiv.org/abs/2203.10097}{{ arXiv:2203.10097~[hep-th]}}.

\bibitem{Debray:2023rlx}
A.~Debray, {\slshape {Bordism for the 2-group symmetries of the heterotic and CHL strings},} \href{http://dx.doi.org/10.1090/conm/802/16079}{{\em Contemp. Math.} {\bfseries 802} (2024) 227--98}, \href{http://arxiv.org/abs/2304.14764}{{ arXiv:2304.14764~[math.AT]}}.

\bibitem{Anber:2024gis}
M.~M. Anber and S.~Y.~L. Chan, {\slshape {Global aspects of 3-form gauge theory: implications for axion-Yang-Mills systems},} \href{http://dx.doi.org/10.1007/JHEP10(2024)113}{{\em JHEP} {\bfseries 10} (2024) 113}, \href{http://arxiv.org/abs/2407.03416}{{ arXiv:2407.03416~[hep-th]}}.

\bibitem{Garcia-Etxebarria:2017crf}
I.~Garc{\'\i}a-Etxebarria, H.~Hayashi, K.~Ohmori, Y.~Tachikawa, and K.~Yonekura, {\slshape {8d gauge anomalies and the topological Green-Schwarz mechanism},} \href{http://dx.doi.org/10.1007/JHEP11(2017)177}{{\em JHEP} {\bfseries 11} (2017) 177}, \href{http://arxiv.org/abs/1710.04218}{{ arXiv:1710.04218~[hep-th]}}.

\bibitem{Freed:2019sco}
D.~S. Freed and M.~J. Hopkins, {\slshape {Consistency of M-Theory on Non-Orientable Manifolds},} \href{http://dx.doi.org/10.1093/qmath/haab007}{{\em Quart. J. Math. Oxford Ser.} {\bfseries 72} (2021) 603--671}, \href{http://arxiv.org/abs/1908.09916}{{ arXiv:1908.09916~[hep-th]}}.

\bibitem{Lee:2022spd}
Y.~Lee and K.~Yonekura, {\slshape {Global anomalies in 8d supergravity},} \href{http://dx.doi.org/10.1007/JHEP07(2022)125}{{\em JHEP} {\bfseries 07} (2022) 125}, \href{http://arxiv.org/abs/2203.12631}{{ arXiv:2203.12631~[hep-th]}}.

\bibitem{Tachikawa:2024ucm}
Y.~Tachikawa and H.~Y. Zhang, {\slshape {On a $\mathbb{Z}_3$-valued discrete topological term in 10d heterotic string theories},} \href{http://dx.doi.org/10.21468/SciPostPhys.17.3.077}{{\em SciPost Phys.} {\bfseries 17} (2024) 077}, \href{http://arxiv.org/abs/2403.08861}{{ arXiv:2403.08861~[hep-th]}}.

\bibitem{Braeger:2025kra}
N.~Braeger, A.~Debray, M.~Dierigl, J.~J. Heckman, and M.~Montero, {\slshape {Cobordism Utopia: U-Dualities, Bordisms, and the Swampland},} \href{http://arxiv.org/abs/2505.15885}{{ arXiv:2505.15885~[hep-th]}}.

\bibitem{Tachikawa:2025flw}
Y.~Tachikawa and K.~Yonekura, {\slshape {On invariants of two-dimensional minimally supersymmetric field theories},} \href{http://arxiv.org/abs/2508.04916}{{ arXiv:2508.04916~[hep-th]}}.

\bibitem{Chakrabhavi:2025bfi}
V.~Chakrabhavi, A.~Debray, M.~Dierigl, and J.~J. Heckman, {\slshape {Exploring Pintopia: Reflection Branes, Bordisms, and U-Dualities},} \href{http://arxiv.org/abs/2509.03573}{{ arXiv:2509.03573~[hep-th]}}.

\bibitem{Dixon:1986iz}
L.~J. Dixon and J.~A. Harvey, {\slshape {String Theories in Ten-Dimensions Without Space-Time Supersymmetry},} \href{http://dx.doi.org/10.1016/0550-3213(86)90619-X}{{\em Nucl. Phys. B} {\bfseries 274} (1986) 93--105}.

\bibitem{Bergman:1999km}
O.~Bergman and M.~R. Gaberdiel, {\slshape {Dualities of type 0 strings},} \href{http://dx.doi.org/10.1088/1126-6708/1999/07/022}{{\em JHEP} {\bfseries 07} (1999) 022}, \href{http://arxiv.org/abs/hep-th/9906055}{{ arXiv:hep-th/9906055}}.

\bibitem{Gaiotto:2017yup}
D.~Gaiotto, A.~Kapustin, Z.~Komargodski, and N.~Seiberg, {\slshape {Theta, Time Reversal, and Temperature},} \href{http://dx.doi.org/10.1007/JHEP05(2017)091}{{\em JHEP} {\bfseries 05} (2017) 091}, \href{http://arxiv.org/abs/1703.00501}{{ arXiv:1703.00501~[hep-th]}}.

\end{thebibliography}\endgroup
\end{document}